\documentclass[a4paper,11pt]{article}
\usepackage{wrapfig}
\pdfoutput=1

\usepackage{jinstpub} 
\usepackage{lineno}

\title{\boldmath Gas electron tracking detector for beta decay experiments}

\author[a,1]{D. Rozpedzik,\note{Corresponding author.}}
\author[b]{L. De Keukeleere,}
\author[a]{K. Bodek,}
\author[b,c,d]{L. Hayen}
\author[a]{K. Lojek,}
\author[a,b]{M.~Perkowski,}
\author[b]{N.~Severijns}

\affiliation[a]{M. Smoluchowski Institute of Physics, Jagiellonian University, PL-30348 Cracow, Poland}
\affiliation[b]{Instituut voor Kern- en Stralingsfysica,
KU Leuven, Celestijnenlaan 200D, B-3001 Leuven, Belgium}
\affiliation[c]{Department of Physics, North Carolina State University, Raleigh, North Carolina 27695, USA}
\affiliation[d]{Triangle Universities Nuclear Laboratory, Durham, North Carolina 27708, USA}

\emailAdd{dagmara.rozpedzik@uj.edu.pl}

\abstract{For identification and 3D-tracking of low-energy electrons a new type of gas-based detector was designed that minimizes scattering and energy loss. The current version of the detector is a combination of a plastic scintillator, serving as a trigger source and energy detector, and a hexagonally structured multi-wire drift chamber (MWDC), filled with a mixture of helium and isobutane gas. The drift time information is used to track particles in the plane perpendicular to the wires, while a charge division technique provides spatial information along the wires. The gas tracker was successfully used in the miniBETA project as a beta spectrometer for a measurement of the weak magnetism form factor in nuclear beta decay. The precision of the three-dimensional electron tracking, in combination with low-mass, low-Z materials and identification of backscattering from scintillator, facilitated a reduction of the main systematics effects. The results originate from performance studies with cosmic muons and low-energy electrons (< 2 MeV) conducted for several pressures (300 – 700 mbar) and isobutane content in the gas mixture (10 – 50\%). At certain conditions, a spatial resolution better than 0.5 mm was obtained in the plane perpendicular to the wires, while resolutions of about 6 mm were achieved along wires. Thanks to precise tracking information, it is possible to eliminate electrons and other particles not originating from the desired decay with high efficiency. Additionally, using the coincidence between MWDC and scintillator, background from gamma emission typically accompanying radioactive decays, was highly suppressed. An overview of different event topologies is presented together with the tracker’s ability to correctly recognize them. The analysis is supported by Monte Carlo simulations using Geant4 and Garfield++ packages. Finally, the preliminary results from the $^{114}$In spectrum study are presented.}


\begin{document}
\maketitle
\flushbottom

\section{Introduction}
At low energy, searches for new effects beyond the Standard Model (BSM) are realized by the high precision experiments e.g. in nuclear and neutron beta decay. Accurate studies of beta-decays have been exploited in various applications of fundamental physics. Especially, the precision measurements of beta spectrum shape and correlation coefficients are widely known~\cite{1}. The dominating contribution in the systematic uncertainty in the beta spectrum shape measurements comes from incomplete deposit of electron energy in the detectors due to backscattering, partial transmission and bremsstrahlung. Monte Carlo simulation of these effects is helpful, however, it introduces its own uncertainty as the input parameters are known with limited accuracy. As the required experimental precision increases, QCD-induced form factors have to be taken into account in order not to limit the sensitivity to BSM physics. 
The BSM parameter called the Fierz interference term $b_\mathrm{F}$ and the Standard Model weak magnetism term $b_{\mathrm{WM}}$ modify the shape of the $\beta$ spectrum for an allowed Gamow-Teller decay as follows \cite{2}:
\begin{equation}
N(W)dW \propto \frac{F(\pm Z, W)}{2\pi^3} pW(W_0-W)^2 \left(1 + \frac{ m_e}{W} b_\mathrm{F}\pm \frac{4}{3}\frac{W}{M_n}\frac{b_{\mathrm{WM}}}{Ac}\right)dW, \nonumber
\end{equation}
where $F(\pm Z, W)$, $p$, $W$, $W_0$ are the Fermi function, the $\beta$ particle momentum, its total energy and total energy at the spectrum endpoint, respectively. Additionally, $m_e$ the electron mass, $M_n$ the nucleon mass, $A$ the nuclear mass number, and $b_{\mathrm{WM}}/c$ the ratio of weak magnetism and Gamow-Teller form factors in the well known Holstein formalism \cite{3}, respectively.

While the Fierz coefficient $b_\mathrm{F}$  contribution is inversely proportional to the total electron energy $W$, the recoil-order terms also affect the spectrum shape with their main contribution, from weak magnetism ($b_{\mathrm{WM}}$), being proportional to $W$~\cite{1,2}. The weak magnetism is a part of the SM but it is still poorly known and new measurements of this quantity with accuracy of about $10\%$ or better are welcome. The sensitivity of the beta spectrum shape to the Fierz and weak magnetism terms was discussed in~\cite{4}. For extraction of the WM form factor from the shape of the beta spectrum, we developed a position sensitive detector that allows for identification and 3D-tracking of electrons while maintaining minimal electron energy~losses.
\newpage
 \section{Multiwire gas electron tracking detector}
 \begin{wrapfigure}{r}{0.6\textwidth}
 \vspace{-10pt}
\centering
\includegraphics[width=.3\textwidth,origin=c,angle=0]{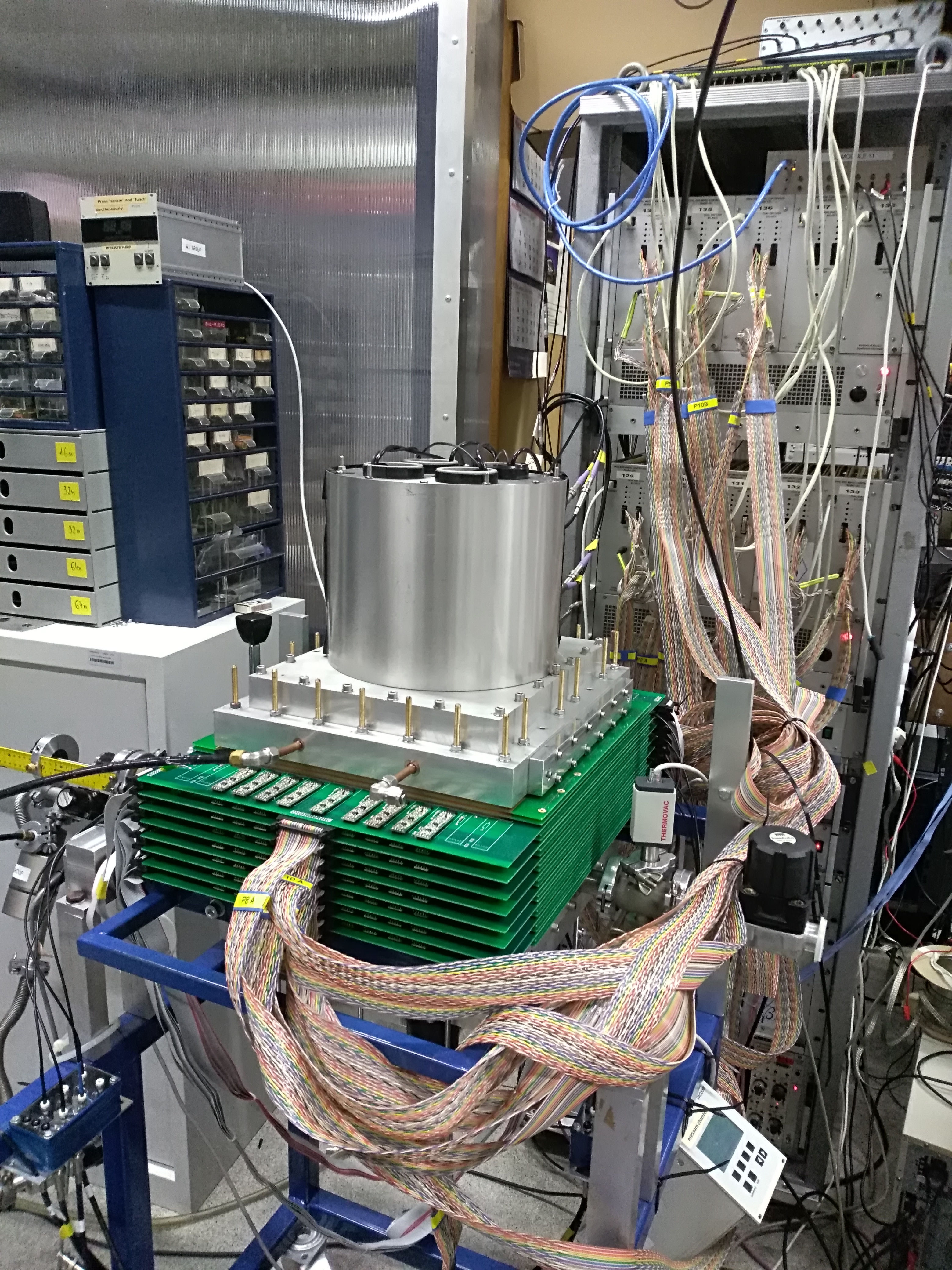}
\includegraphics[height=61mm,width=.25\textwidth,origin=c,angle=0]{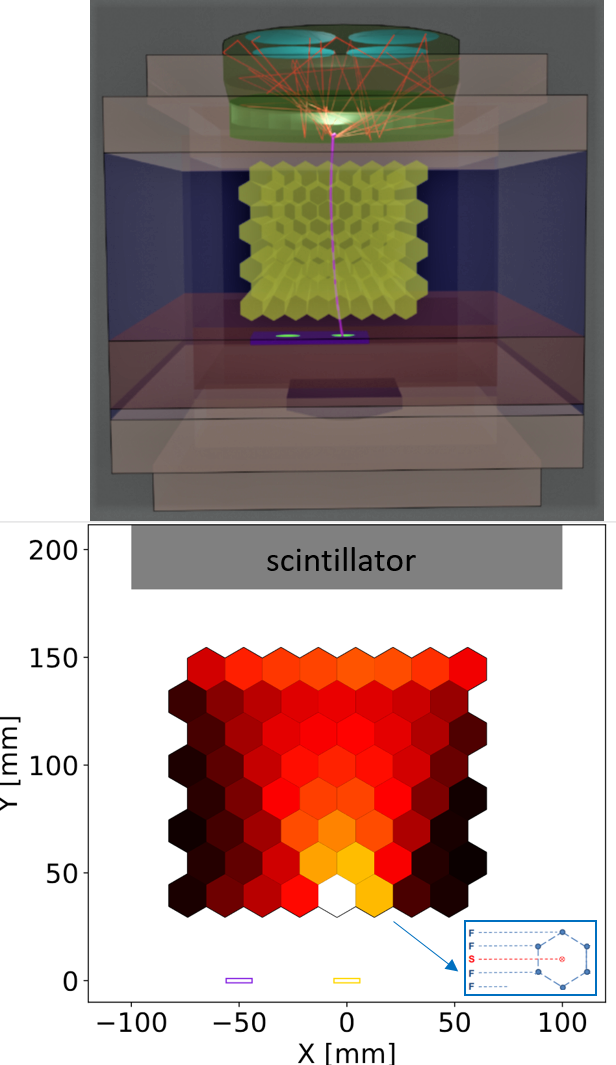}
\caption{\label{fig:1}The miniBETA spectrometer. Left panel: Photo of the setup. Inside view of the chamber as implemented in MC simulation. In the right-lower panel the cells colors indicate intensity of hits. Electrons are emitted from the beta sources located in the middle of the chamber.}
\end{wrapfigure}
The multiwire gas electron tracking detector named miniBETA spectrometer was built for studying experimental effects that must be controlled in $\beta$ spectrum shape measurements. The current version of miniBETA is a combination of a plastic scintillator (serving as energy detector and a trigger source) and a hexagonally structured multi-wire drift chamber (MWDC), filled with a mixture of helium and isobutane at low pressure. The gas electron tracker is responsible for efficient identification of electrons emitted from the $\beta$ decay source. Thanks to precise information about the electron track, it is possible to identify electrons backscattered from the energy detector and eliminate electrons not originating in the $\beta$ source. Additionally, the coincidence condition between signals from the tracker and energy detector suppresses background from gamma emission typically accompanying $\beta$ decays. The light construction of the  MWDC and optimized geometry help reducing background from secondary radiation created inside the chamber due to collisions with wires and mechanical support structure. The hexagonal cell configuration was chosen to assure maximum transparency of the detector in order to minimize multiple electron scattering on wires. The present MWDC configuration, optimal for an overall performance characterization, consists of 8 sense wire planes sandwiched between 24 field wire planes forming the plane structure shown in Fig.~\ref{fig:1}. Each cell consists of a thin anode wire (NiCr alloy, 25 $\mu$m thick) surrounded by 6 cathode field wires (CuBe, 75 $\mu$m thick) forming a hexagonal cuboid. The choice of the anode wire material is dictated by the charge division technique delivering information about the hit position coordinate along wires. The inside view of MWDC is shown in Fig.~\ref{fig:1} (right upper panel) with the illumination of cells from a real measurement~(right~lower~panel). 

The energy detector is made of a plastic scintillator embedded in the gas detector and connected via a lightguide with four photomultiplier tubes (PMT) installed outside the chamber. The digitized pulse height of PMT signals carry the electron energy information. 
\begin{figure}[t]
\vspace{-10pt}
\centering
\includegraphics[width=\textwidth,origin=c,angle=0]{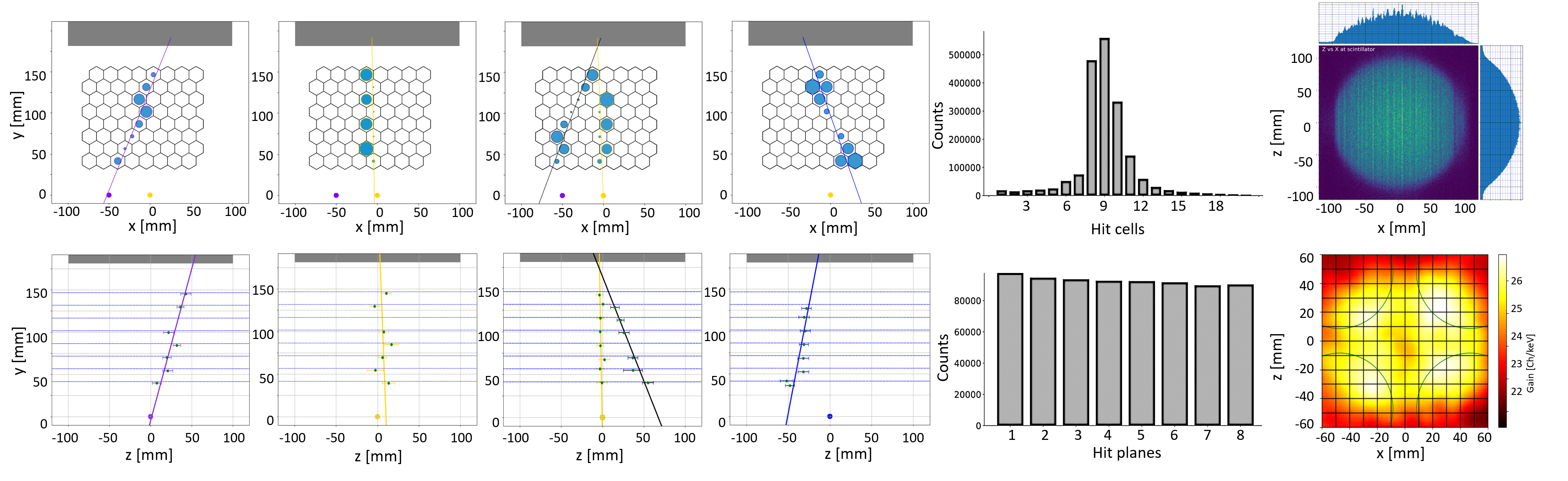}
\caption{\label{fig:2} From Left: Different event topologies recognized by the 3D-tracking algorithm: ${^{207}}$Bi, ${^{114}}$In, backscattered and cosmic events, respectively. Next, the hit and plane multiplicity distributions. The right upper plot shows the 2d-distribution of fitted trajectories extrapolated to the scintillator surface with projection along $z$ an $x$ directions. The right bottom plot shows the energy detector gain map.}
\end{figure}
Additionally, the PMT signals provide time reference for drift time measurement and are also used as a trigger for the readout sequence including MWDC data and information from the electron energy detectors. The miniBETA spectrometer was operated with He/isobutane gas mixtures ranging from 50/50$\%$ to 90/10$\%$ at pressures from 700 down to 300 mbar. The spectrometer is read out by a custom-designed modular electronic system described in Refs. \cite{5,6,7}. 
The 3D tracking algorithm extracts the XY- coordinates of points on the electron track using the information from drift time and ZY- coordinates from the charge division method. Examples of the recorded event topologies are presented in Fig. \ref{fig:2}.
An average efficiency of the trajectory identification greater than 98$\%$ was reached for most of the cells. The spatial resolution of 3D-tracking varies between 0.2 and 0.8 mm for $(x,y)$ and 4 -- 10 mm for $z$ coordinate of a hit in a cell. It depends on the track distance from the signal wire and on the signal amplitude for a given gas mixture~\cite{6,7}.
\vspace{-4pt}
\section{Analysis results of measured $\beta$ spectra shapes}
\begin{figure}[b]
\includegraphics[width=\textwidth,origin=c,angle=0]{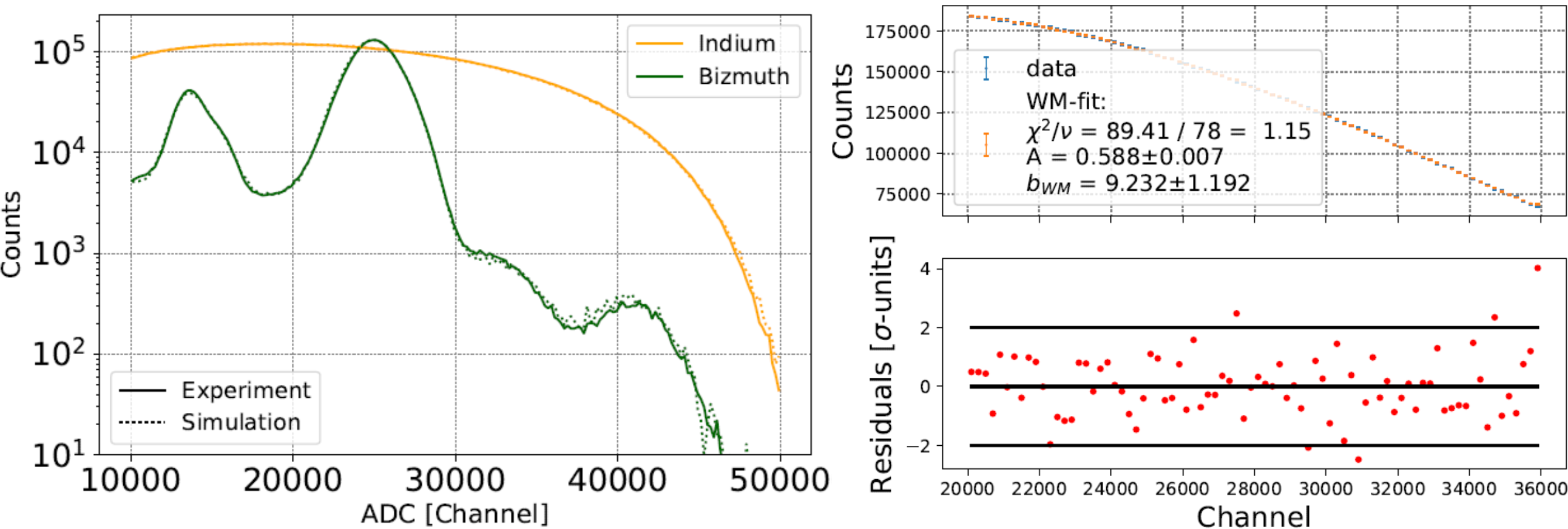}
\caption{\label{fig:4} Left panel: A comparison of the recorded experimental and simulated spectra of $^{207}$Bi conversion electrons and ${^{114}}$In. Right panel: The preliminary results for $b_{\mathrm{WM}}$ extracted from ${^{114}}$In spectrum.}
\end{figure}
The presented experimental setup was fully modelled in MC simulations ~\cite{7}. The total experimental and simulated $\beta$ spectra of $^{207}$Bi and $^{114}$In (assuming $b_\mathrm{F}$ and $b_{\mathrm{WM}}$ to vanish) are shown in Fig.~\ref{fig:4}. The experimental spectra were found to be reproduced by simulation at the $10^{-2}$ level. In the ${^{207}}$Bi spectrum, the corresponding peaks of the measured conversion electrons from K, L and M shells are well matched to the simulation. The comparison of the measured and simulated ${^{114}}$In spectra exhibit a slope difference on $10^{-2}$ level in the energy region 700 -- 1600 keV, which could be due to $b_{\mathrm{WM}}$ term. Using the same simulation parameters, the detector response was simulated and the weak magnetism form factor was extracted by fitting to the theoretical $\beta$ spectrum shape~\cite{2} for varying $b_{\mathrm{WM}}$ values. The preliminary result of this procedure reveals a $b_{\mathrm{WM}} = 9.2 \pm 1.2\,(\mathrm{stat})$. The systematic error analysis is currently ongoing.

\section{Summary and outlook}
Application of the 3D gas electron tracker with a plastic scintillator for beta spectrum shape measurements gives promising results. The measurements of the spectrum shape for ${^{114}}$In and ${^{32}}$P isotopes are completed. The reached sensitivity of the electron track reconstruction and energy resolution supported by very good description by MC simulations allowed the extraction of the weak magnetism term from ${^{114}}$In $\beta$ spectrum shape. The analysis of the ${^{32}}$P spectrum is in progress. 

\begin{small}

\end{small}

\end{document}